\documentclass[multphys,vecphys]{svmult}

\usepackage{makeidx}         % allows index generation
\usepackage{graphicx}        % standard LaTeX graphics tool
\graphicspath{{fig/}}
\usepackage{multicol}        % used for the two-column index
\usepackage[bottom]{footmisc}% places footnotes at page bottom
\makeindex             % used for the subject index
\begin{document}

\title*{The XENON Dark Matter Experiment}
\author{Elena Aprile (on behalf of the XENON collaboration)}
\institute{Physics Department and Columbia Astrophysics Laboratory, Columbia University, New York, New York 10027\\
\texttt{age@astro.columbia.edu}}
\maketitle

The XENON experiment aims at the direct detection of dark
matter in the form of WIMPs (Weakly Interacting Massive Particles) via their
elastic scattering off Xenon nuclei. With a fiducial mass of 1000 kg of liquid xenon, a sufficiently low threshold of ~16 keV recoil energy and an un-rejected background rate of 10 events per year, XENON would be sensitive to a WIMP-nucleon interaction cross section of $\sim 10^{-46} \rm cm^2$,  for WIMPs with masses above 50 GeV.   
The 1 tonne scale experiment (XENON1T) will be realized with an array of ten identical 100 kg detector modules (XENON100). The detectors are  time projection chambers operated
in dual (liquid/gas) phase, to detect simultaneously the ionization, through secondary
scintillation in the gas, and primary scintillation in the liquid produced by low energy recoils. The
distinct ratio of primary to secondary scintillation for nuclear
recoils from WIMPs (or neutrons), and for electron recoils from background, is
key to the event-by-event discrimination capability of XENON. A 3kg dual phase detector with light readout provided by an array of 7 photomultipliers is currently being tested, along with other
prototypes dedicated to various measurements relevant to the XENON program. We present some of the results obtained to-date and briefly discuss the next step in the phased approach to the XENON experiment, i.e. the development and underground deployment of a 10 kg detector (XENON10) within 2005. 
\section{Introduction}
\label{intro}

The question of the nature of the dark matter in the Universe is being addressed with numerous direct and indirect detection experiments using a variety of methods, detectors and target materials. For a recent review of the field we refer to \cite{Gaitskell:05}. 

The proposed XENON experiment is among the new generation direct searches for dark matter weakly interacting massive particles (WIMPs) with the ambitious goal of a sensitivity reach which is several orders of magnitude higher than the lowest exclusion limit set by the CDMS II experiment\cite{CDMS:04} 
To achieve a sensitivity goal of $\sim$
10$^{-46}$ cm$^2$ XENON relies on a target mass of 1 tonne of liquid xenon (LXe), with less than about 10 background events per year. Efficient background identification and reduction is based on the distinct ratio of the ionization and scintillation signals produced in LXe by nuclear (from WIMPs and neutrons)  and electron (from gamma, beta and alpha backgrounds) recoil events \cite{Yamashita:03}. The main challenge is to accomplish this event-by-event discrimination down to a few tens of keV nuclear recoil energy. 
Additional techniques used for background suppression are an active LXe self shield around the sensitive target, passive gamma and  neutron shielding and the detector's 3-D position resolution.  The position information is crucial to select single hit events characteristic of a WIMP signal and to veto multiple hit events associated with neutrons as well as other backgrounds which propagate from the edge of the detector into the fiducial volume.

To test the XENON concept and verify achievable threshold, background rejection power and sensitivity, a detector with a fiducial mass on the order of 10 kg (XENON10), is under development for underground deployment in 2005. The detector exploits several key systems which have been tested and optimized with the 3 kg prototype, but will feature significant improvement in overall performance and sensitivity down to 16 keV nuclear recoil energy.  The experiment will be carried out  at the Gran Sasso Underground  Laboratory  (3500 mwe). The depth and the expected background rejection power  will allow us to reach a sensitivity a factor of 20 below the best existing measurements from CDMS II \cite{CDMS:04}, of ~2 dark matter events/10 kg/month, without the need of a muon veto for fast neutrons. 

Another important goal of the XENON10 phase is to pave the way for the design of a 100 kg scale detector. With 3 months of operation deep underground, at a background level below $1 \times 10^{-5}$ cts/keVee/kg/day after rejection, XENON100 would provide a sensitivity of $\sim 10^{-45} \rm cm^2$.  The full 1 tonne scale experiment (XENON1T) will be realized with ten XENON100 modules.

\begin{figure}[htbp] 
  \begin{center} 
   \includegraphics[width=7cm]{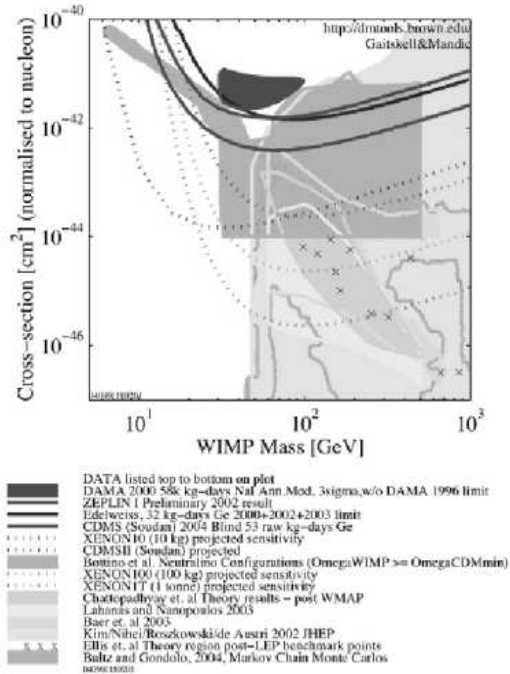} 
   \caption{Theoretically predicted regions for SUSY WIMP candidate, along with the best detection dark matter limits from current direct detection experiments. Also shown as four dotted lines (top to bottom at right) are the projected sensitivities for CDMS II at Soudan \cite{CDMS:04}, and for XENON10, XENON100 and XENON1T \cite{NSF:01}.}   
   \label{sensitivity} 
  \end{center} 
\end{figure}

Fig. \ref{sensitivity} shows the sensitivity projected for XENON10 experiment, in comparison to current WIMP searches, which are probing event rates at $\sim$ 0.1 evts/kg/day. The projected performance of XENON100 and XENON1T detectors is also shown. In order to continue progress in dark matter sensitivity it will be important to have a liquid xenon experiment at the 10 kg scale operational and taking science data in early 2006.

\section{The XENON Detector Baseline}
\begin{figure}[htbp] 
  \begin{center} 
   \includegraphics[width=5cm]{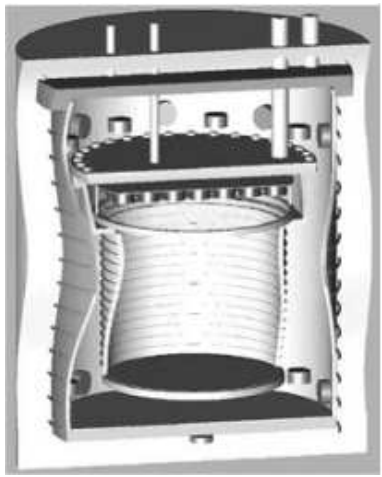} 
   \caption{Schematic view of the XENON 100 dual phase detector.}   
   \label{xe100} 
  \end{center} 
\end{figure}

Fig. \ref{xe100} schematically shows the design of the detector proposed as unit
module for the XENON experiment. It is a dual phase TPC, with the active LXe
volume defined by a 50 cm diameter CsI photocathode immersed in the
liquid, at about 30 cm from the first of three wire grids defining a gas
proportional scintillation region. An array of compact, metal channel UV sensitive PMTs developed by Hamamatsu Photonics Co. to work at LXe temperature and recently optimized for low radioactivity content, are used for primary and secondary light detection. 

The TPC is enclosed in a leak-tight cylindrical structure made of
PTFE and OFHC. The PTFE is used as effective UV light reflector \cite{Yamashita:04}
and as electrical insulator. The fraction of direct light heading
downward will be efficiently detected with the CsI photocathode
\cite{CsI}. The whole structure is immersed in a bath of LXe, serving
as active veto shield against background. The LXe for shielding
is readout by PMTs.

An event in the XENON TPC will be characterized by three signals
corresponding to detection of direct scintillation light, proportional
light from ionization electrons and CsI photoelectrons. Since electron diffusion in LXe is
small, the proportional scintillation pulse is produced in a small spot
with the same X-Y coordinates as the interaction site, allowing 2D
localization with an accuracy of 1 cm. With the more precise Z
information from the drift time measurement, the 3D  event localization
provides an additional background discrimination via fiducial volume
cuts. The simulated detection efficiency of the primary scintillation light is
about 5 p.e./keV for the XENON100 detector. \
\section{Results from the 3kg XENON prototype}
\label{}
R\&D for the XENON program is being carried out with various prototypes dedicated to test several feasibility aspects of the proposed concept, and to measure the relevant detector characteristics such energy threshold and background discrimination as well as ionization and scintillation efficiency of Xe recoils in LXe as a function of energy and electric field. Here we limit the discussion to the results obtained to-date with a dual phase xenon prototype with $\sim$ 3 kg of active mass. The primary 
scintillation light ($\rm S1$) from the liquid, and the secondary scintillation
light ($\rm S2$) from the ionization electrons extracted into
the gas phase, are detected  by an array of seven PMTs, operating in the cold gas above the liquid. 

\begin{figure}[htbp] 
  \begin{center} 
   \includegraphics[width=5cm]{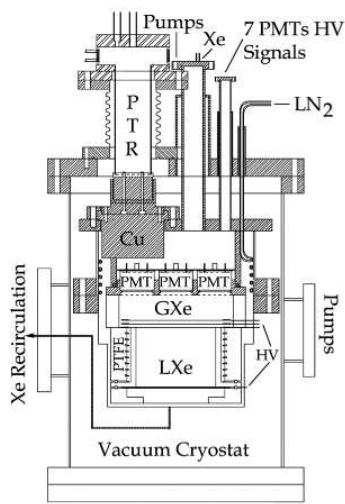} 
   \caption{Schematic drawing of the dual phase prototype.}   
   \label{prototype} 
  \end{center} 
\end{figure}

\begin{figure}[htbp] 
  \begin{center} 
   \includegraphics[width=6cm]{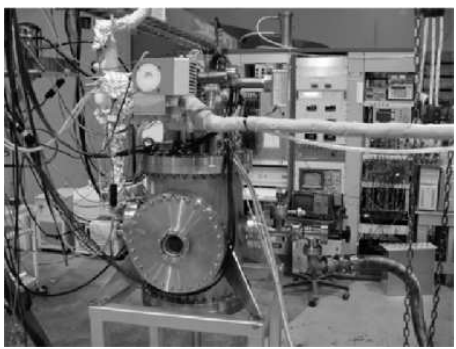} 
   \caption{The detector integrated with the vacuum cryostat, refrigerator, gas/recirculation and DAQ systems.}   
   \label{setup} 
  \end{center} 
\end{figure}

A drawing of the 3kg XENON prototype is shown in Fig. \ref{prototype} while
the photo of Fig. \ref{setup} shows the integrated set-up at the Columbia University Nevis Laboratory. The sensitive volume of the TPC (7.7$\times$7.7$\times$5.0 cm$^3$) is
defined by PTFE walls and grids with high optical transmission, made of Be-Cu wires with a pitch of 2 mm
and 120$\mu$m diameter. Negative HV is applied to the bottom grid, used as cathode. Grids on the top close the charge drift region in the liquid
and with appropriate biasing, create the amplification region for gas proportional scintillation.
Shaping rings located outside of the PTFE
walls and spaced 1.5 cm apart, are used to create a uniform electric field for
charge drift.

\begin{figure}[] 
  \begin{center} 
   \includegraphics[width=5cm]{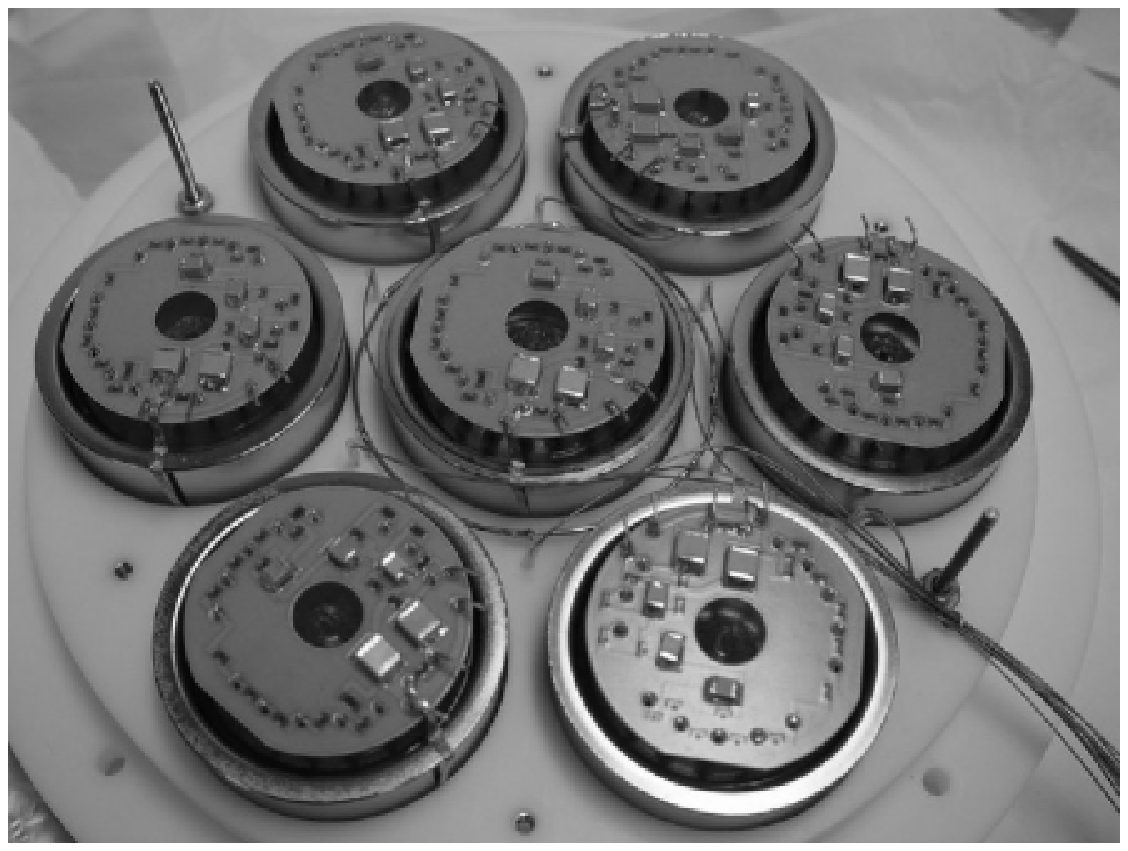} 
   \caption{An array of 7 PMTs on the top of the chamber in the gas phase.}   
   \label{pmt_array} 
  \end{center} 
\end{figure} 

An array of seven, 2 inch PMTs (Hamamatsu R9288), mounted 2.3 cm above the top grid, is used  to detect both  primary and proportional light . The PMT array mounted on a PTFE frame is shown in Fig. \ref{pmt_array}. The custom-developed HV divider bases are also clearly visible. We used LEDs to measure the PMTs gain and single photoelectron response.

The LXe detector is insulated by a vacuum cryostat. A Pulse Tube Refrigerator (PTR) optimized for LXe, is used to cool down the detector, liquefy Xe gas and maintain the liquid temperature at the desired value. The typical operating temperature is -100 $\rm ^o$C with a stability better than 0.05 $\rm ^o$C. At this temperature the Xe vapor pressure is $\sim$1.8 atm. A reliable and stable cryogenics system is an essential requirement for the XENON experiment since both PMT's gain and the proportional light yield vary with temperature.  With a cooling capacity of 100 W at 165 K the  same PTR equipment will be used for XENON10 underground. 
	
\begin{figure}[htbp] 
  \begin{center} 
   \includegraphics[width=6cm]{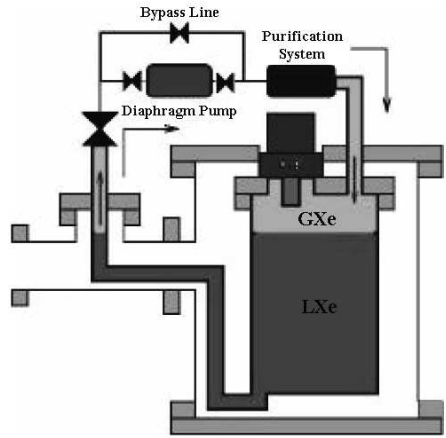} 
   \caption{Schematics of the continuous Xe circulation and purification system.}   
   \label{recir_sys} 
  \end{center} 
\end{figure}

The XENON experiment requires ultra high purity LXe to enable ionization electrons to drift freely over the 30 cm proposed for the XENON100 unit module. Furthermore, the LXe purity has to be maintained during the long-term detector operation required for statistics and  annual modulation analysis. A Xe purification, re-circulation and recovery system was built and operated with the 3 kg prototype \cite{Mihara:04}(see Fig. \ref{recir_sys}).

For Xe gas purification, a single high temperature SAES getter was used\cite{SAES}. Electron lifetime longer than 500 $\mu$sec is routinely achieved after a few days of continuous purification. A similar gas system with the addition of a Kr removal section will be used for XENON10 underground.  We plan to start with commercial Xe gas with a Kr level of roughly 10 ppb to be reduced to a level well below 1 ppb by an adsorption-based system currently under construction.  The reliability and efficiency of both the cryogenics and gas purification systems have been tested with repeated experiments lasting for several weeks continuously.

\begin{figure}[htbp] 
  \begin{center} 
   \includegraphics[width=10cm]{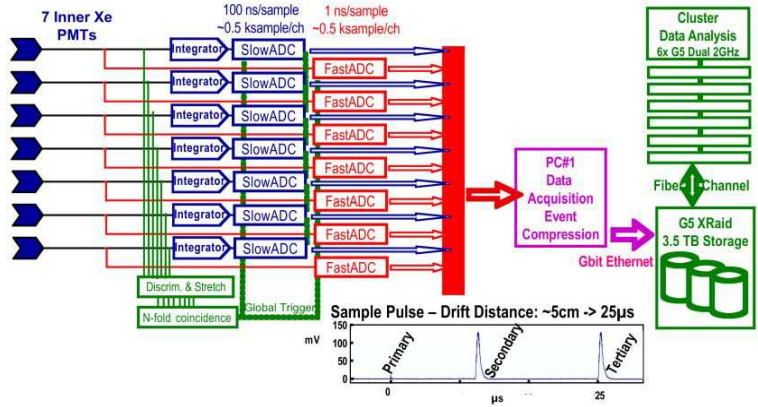} 
   \caption{The DAQ system for recording S1, S2 and S3 signals from seven PMTs.}   
   \label{DAQ} 
  \end{center} 
\end{figure}

To maximize the information available from time structure and amplitude of the primary (S1), secondary (S2) and possible CsI photoelectron signal (S3), the 7 PMTs are digitized by both fast (1 ns/sample, 8 bit) and slow (100 ns/sample, 12 bit) ADCs. The gain of the fast ADCs is matched to observe signals down to single photoelectrons, whereas the slow ADCs are optimized to observe the longer proportional light signals from S2/S3. The DAQ system  has been developed and successfully applied to record S1, S2 and S3 signals from the current prototype (Fig. \ref{DAQ}). The coincidence of more than one PMT signals was required to create a trigger. 

\begin{figure}[htbp] 
  \begin{center} 
   \includegraphics[width=9cm]{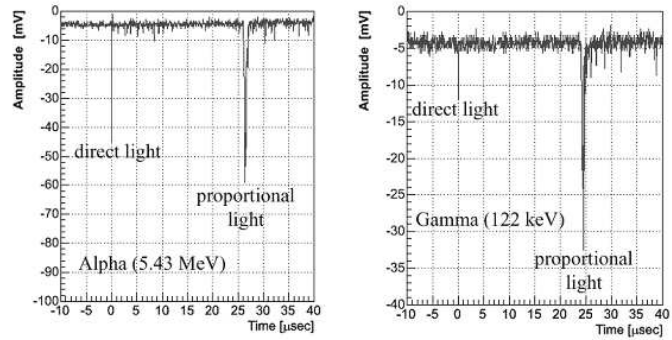} 
   \caption{Waveforms of direct and proportional light for an alpha recoil (left) and an electron recoil (right).}   
   \label{ag_waveform} 
  \end{center} 
\end{figure}

The 3kg dual phase detector's operation  was tested using low energy gamma-rays from a Co-57 source, alpha particles from Po-210 deposited on the cathode, and neutrons from an AmBe source.  Two typical waveforms of the direct light produced in the liquid (S1) and the proportional light produced by electrons extracted in the gas (S2) are shown in Fig.\ref{ag_waveform}, for an alpha and gamma event. The S1 signals are prompt while the S2 signals have a width of a few  $\mu$sec, as expected.   The time separation between S1 and S2 is dominated by the drift time in the liquid so that the position of the source along the drift axis is accurately inferred from the known drift velocity at the applied electric field.  Since both the Co-57 122 keV gammas and the Po-210 5.43 MeV alphas are very localized in the dense LXe, the time separation between S1 and S2 is close to the maximum drift time of 25 $\mu$sec. The other two coordinates are inferred from the center of gravity of the proportional light emitted near the 7 PMTs.   Simulations and a preliminary analysis of the alpha data show that the transverse direction of the source can be localized with an accuracy of 1 cm.

\begin{figure}[htbp] 
  \begin{center} 
   \includegraphics[width=9cm]{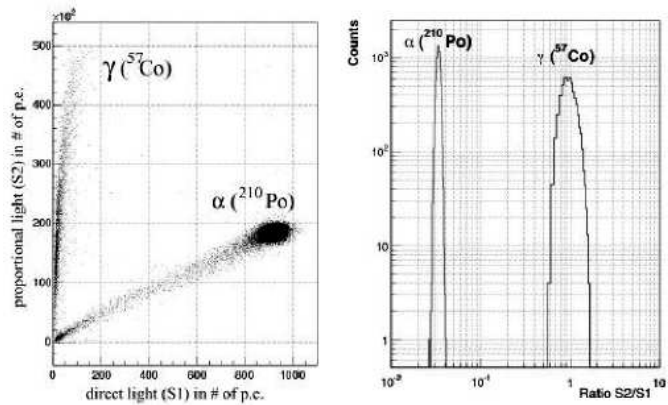} 
   \caption{Measured distribution of  S2 (proportional light) versus S1 (direct light) for combined alpha and gamma events (left).
Distribution of alpha and gamma events plotted as a function of the ratio S2/S1 (right).}
   \label{fig:ratio} 
  \end{center} 
\end{figure}

In Fig. \ref{fig:ratio}(left) the S1 and S2 signals, simultaneously recorded with gamma and alpha irradiation, are plotted together. The detector was operated at 1kV/cm in the drift region and
10 kV/cm in the gas region. The two classes of events are well differentiated. Another visualization of this event separation is shown in Fig. \ref{fig:ratio} (right) where the distribution of the same events is plotted as a function of the S2/S1 ratio, in logarithmic scale.  The peak from gamma rays is normalized to 1. 

The measured ratio $\rm (S2/S1)_{\alpha}/(S2/S1)_{\gamma}$ is about 0.03. The separation between alpha recoils and electron recoils is already remarkable, despite the non optimized light collection of the detector at this stage. The ratio is even larger, if we account for the fraction of primary light produced by alpha recoils which is absorbed by the $^{210}$Po source disk.
  Another distinct feature that separates alpha recoils from electron recoils is the dependence of the light on applied electric field.  While the primary light from an electron event is strongly quenched by the field because of the reduced recombination rate, this is not the case for a heavily ionizing particle such as an alpha. This means that the primary light is barely affected by the field and the S2/S1 ratio is essentially constant. The dependence of the primary light on the applied field was previously measured by Aprile et al. \cite{Aprile:90} and has been verified with data from the XENON prototypes. 

%Add Rick's section from here

The $S2/S1$ ratio in LXe for nuclear recoil events was established
using a 5 Ci $^{241}$AmBe source, emitting neutrons in the energy range
0--8 MeV, in conjunction with
a BC501A scintillation coincidence counter to detect events scattering from the LXe target.
The 1.4 liter BC501A counter was placed at a distance of
50 cm from the center of the LXe chamber, at a scattering angle of 130 $\deg$.

\begin{figure}[htbp] 
  \begin{center} 
   \includegraphics[width=8cm]{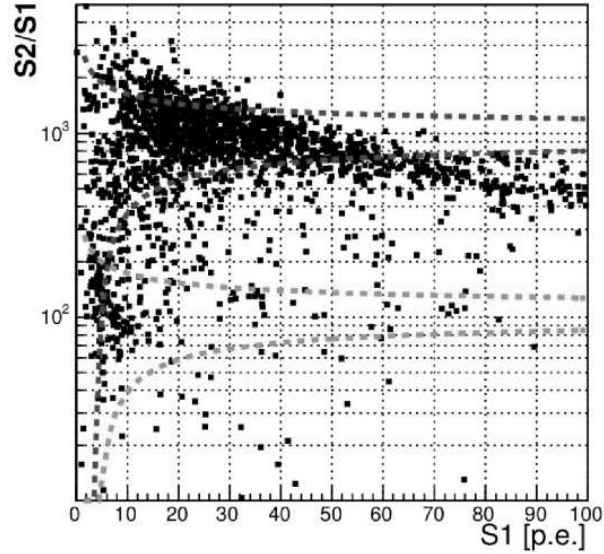} 
   \caption{Events in LXe from an AmBe source, tagged using the
   method described in the text. Events with $S2 >$ 50k photoelectrons 
   are rejected since they saturate the acquisition electronics. The region defined
   by the red dashed lines represents a band with $S2/S1\sim$ 1000 associated with
   electron recoil events. The region defined by the green dashed lines
   represents a band with $S2/S1\sim$ 100 associated with elastic nuclear recoil events.
   Inelastic neutron scattering events are also present. Their predicted event distribution is 
   shown in Fig. \ref{AmBe_sim}.
   }
   \label{3kg_AmBe_S2S1} 
  \end{center} 
\end{figure}

\begin{figure}[htbp] 
  \begin{center} 
   \includegraphics[width=8cm]{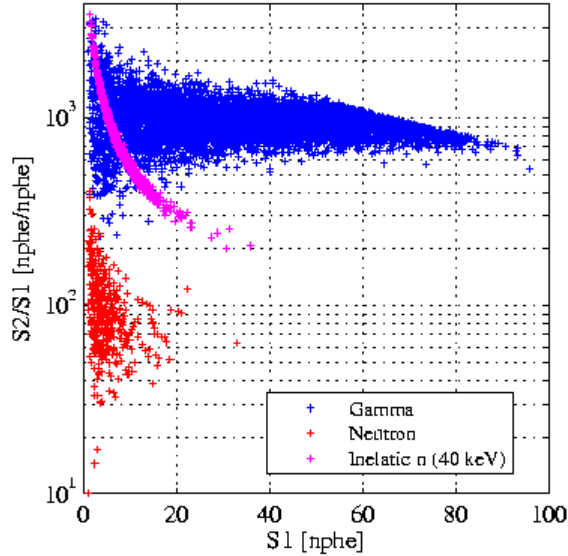} 
   \caption{Simulation of detector response, $S2/S1$ vs $S1$, for AmBe neutrons 
   and uniform gamma spectrum. Distributions are shown for 
   electron recoils (blue, $S2/S1\sim$ 1000), 
   neutron elastic recoils (red, $S2/S1\sim$ 100), 
   and neutron inelastic recoils
   associated with $^{129}$Xe 40 keV excited state. 
   The inelastic scattering for the $^{131}$Xe 80 keV excited state is not shown.
   The statistical fluctuations, 
   associated with all steps in the generation of S1 and S2 signals
   have been considered. 
   The variation in signals with position within the detector is not simulated for this plot.
   }
   \label{AmBe_sim} 
  \end{center} 
\end{figure}

A population of candidate neutron
scattering events in the LXe was obtained by identifying events which 
were
(i) tagged as neutron recoils 
in the BC501A (by pulse shape discrimination) 
and (ii) also had an implied ToF (time of flight)
between the LXe and BC501A in a window of 40--70 ns.
The selected events also contained a significant population 
of accidental coincidences, between gammas  
scattering in the LXe, and 
neutrons, emitted separately, which interact in the BC501A
 in the appropriate time window.
Figure \ref{3kg_AmBe_S2S1} shows LXe event data, for the AmBe source, in which
there is a population of events with $S2/S1\sim 1000$,
associated with electron recoils, 
and a second population of events, with $S2/S1\sim 100$, associated with elastic nuclear recoils. 
The figure also contains events arising from the inelastic scattering of neutrons
from $^{129}$Xe (nat. abun. 27\%)  and $^{131}$Xe (nat. abun. 21\%) which have excited states of 40 keV, and 80 keV, respectively. 
A simulation of the predicted event distribution $S2/S1$ vs. $S1$ is shown in Fig. \ref{AmBe_sim}
for comparison with the data in Fig. \ref{3kg_AmBe_S2S1}.

\begin{figure}[htbp] 
  \begin{center} 
   \includegraphics[width=6cm]{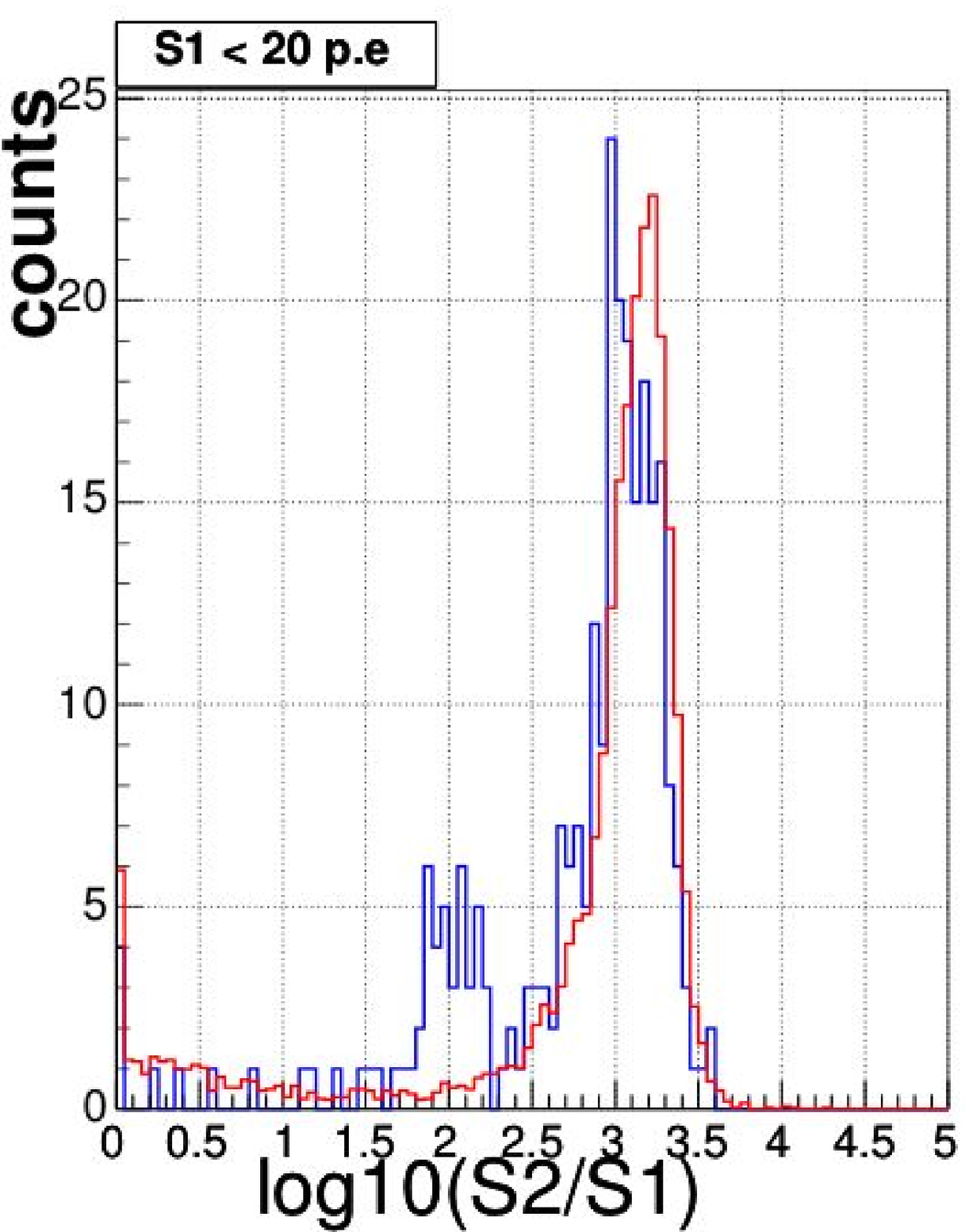} 
   \caption{The blue (dark) line shows a histogram of $S2/S1$ for events
   with $S1<$ 20 p.e. taken for AmBe source data shown in Fig. \ref{3kg_AmBe_S2S1}. 
   For comparison 
   the $S2/S1$ distribution (red (light) line) 
   for Compton electron recoil events, in the same $S1$ range,
   for a $^{137}$Cs source of 662 keV gamma rays is shown. 
   The AmBe curve shows two distinct populations associated with electron and
   neutron recoils at  $S2/S1\sim$ 1000 and  $S2/S1\sim$ 100, respectively.
   The second population is absent in the $^{137}$Cs data. 
   }
   \label{3kg_AmBe_S2S1hist} 
  \end{center} 
\end{figure}

A histogram comparing the $S2/S1$ distributions for events $S1 <$ 20 p.e.
for the AmBe source, and separately a $^{137}$Cs source, 
are shown in Fig. \ref{3kg_AmBe_S2S1hist}.
The AmBe curve shows the two populations associated with electron and
 nuclear recoils.
 The second population is absent in the $^{137}$Cs data. 
 The leakage of electron recoil events
 in the $^{137}$Cs data
 into the $S2/S1$ region for nuclear recoil events
 is $< 1\%$.
It was established, using separate calibration work, that the
 the $S1$ signals for electron and nuclear recoils is 0.14 p.e./keV$_{ee}$ 
 and 0.07 p.e./keV$_{r}$, respectively.

%3kg_AmBe_ToFTagged_From_KNiFig2.eps
% S2/S1 vs S1 plot

% 3kg_S2S1_GammaNeutronInelN_v2.eps
% Simulated S2/S1

% 3kg_AmBe_Hist_From_KNiFig4.eps
% S2/S1 plot

\begin{figure}[htbp] 
  \begin{center} 
   \includegraphics[width=8cm]{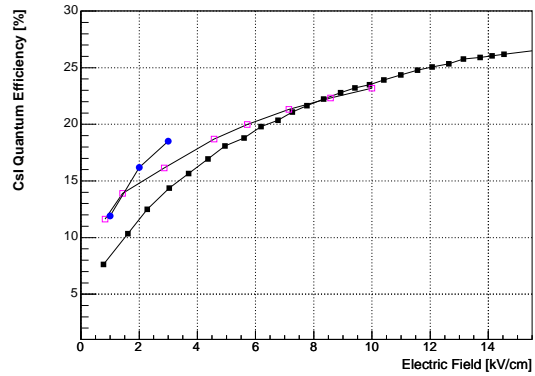} 
   \caption{CsI QE as a function of electron extraction field. The solid square data points are from \cite{CsI}. The open square are from recent measurements with a similar setup as in \cite{CsI}. The circle data points are inferred from the signals measured with the 3kg dual phase prototype.}
   \label{csiqe_e} 
  \end{center} 
\end{figure}

The light collection efficiency of the 3kg prototype is not yet optimized, as a large fraction of UV photons are lost by total internal reflection at the liquid-gas interface. As originally proposed for the XENON baseline detector, a CsI photocathode in place of a common cathode can significantly improve light collection and lower the minimum energy threshold \cite{NSF:01}. Monte Carlo simulations show that the primary light collection efficiency of the 3kg prototype would increase to $\sim$6 p.e./keV, with the cathode grid replaced by a CsI photocathode. Results from recent tests with various photocathodes are very encouraging. We have confirmed the high QE in LXe (see Fig.\ref{csiqe_e}), first measured by the Columbia group more than ten years ago \cite{CsI}. We have also demonstrated the effective suppression of the photon feedback connected with a CsI in a dual phase detector, using a commercial  HV  switch  unit (PVX-4130 from Directed Energy, Inc). The normal rise and fall time of $<$ 100 nsec was slowed to 1 $\mu$sec, and not appreciable noise from the switching was observed on the PMTs in our 3 kg prototype. Proportional scintillation could be stopped as expected by deriving an appropriate gate signal from the light trigger.

From measurements with a CsI photocathode in the 3kg dual phase detector we infer a similar value for QE as that measured with a simple ionization chamber without PMT and switching supply. The QE of the photocathode as a function of field (up to 3 kV/cm) is derived  from a Monte Carlo simulation of the light collection efficiency and the measured signal size ratio between S2 and S4, where S2 is as explained in the text and S4 is the proportional light from the extracted photoelectrons induced by S2 on the CsI photocathode. The photocathodes used in these experiments were deposited at the same time, using the same substrate and thickness of the CsI layer. Combined results are shown in Fig.\ref{csiqe_e}. We are finalizing a CsI deposition apparatus at Columbia which will enable us to optimize preparation and test CsI photocathode for maximum performance in LXe.

The experience gained with the 3kg prototype, its performance and results to-date, as well as results obtained with other detectors not presented here \cite{Aprile:PRD05}, are guiding the realization of XENON10. The 10 kg scale detector will use a light readout with a CsI photocathode in the liquid and an array of PMTs in the gas for much improved light detection efficiency and sensitivity to low energy recoils.

This work was supported by a grant from the National Science Foundation to the Columbia Astrophysics Laboratory (Grant No. PHY-02-01740).
%
%
% BibTeX users please use
% \bibliographystyle{}
% \bibliography{}
%
% Non-BibTeX users please follow the syntax
% the syntax of "referenc.tex" for your own citations
%%%%%%%%%%%%%%%%%%%%%%%% referenc.tex %%%%%%%%%%%%%%%%%%%%%%%%%%%%%%
% sample references
% "physics"
%
% Use this file as a template for your own input.
%
%%%%%%%%%%%%%%%%%%%%%%%% Springer-Verlag %%%%%%%%%%%%%%%%%%%%%%%%%%

%
% BibTeX users please use
% \bibliographystyle{}
% \bibliography{}

%%%%%%%%%%%%%%%%%%%%%%%%%%%%%%%%%%%%%%%%%%%%%%%%%%%%%%%%%%%%%%%%%%%%%%  }

%%%%%%%%%%%%%%%%%%%%%%%%%%%%%%%%%%%%%%%%%%%%%%%%%%%%%%%%%%%%%%%%%%%%%%

\printindex
\end{document}